\documentstyle[aps,prl]{revtex}
\begin{document}
\author{X.X. Yi$^{1,2}$,\ \ D.L.Zhou$^2$,\ \ \ C.P.Sun$^2$\ \ ,W.M.Zheng$^2$ }
\address{ $^1$ Institute of Theoretical Physics, Northeast Normal
University, Changchun 130024, China\\
$^2$Institute of Theoretical Physics, Academia Sinica, P.O.Box 2735, Beijing 100080, China}
\title{Finite Temperature Excitations of an Inhomogeneous
Trapped Bose Gas With Feshbach Resonances{\footnote{Mailing address: Institute of Theoretical
Physics, Academia Sinica, Peking 100080, China}}
}
\maketitle
\begin{abstract}
We derive and discuss the temperature dependance of the condensate and
noncondensate density profile of a Bose-Einstein condensate gas with Feshbach
resonance in a parabolic trap. These quantities are calculated
self-consistently using the generalized Hartree-Fock-Bogoliubov (HFB)
equations
within the Bogoliubov approximation.
At zero temperature, the HFB equation can be solved by means of
a variation method that give the low excitation spectrum.
Moreover, within the two-body collision theory, we estimate the
relationship between the atom number and the external magnetic field
$B$, it is in good agreement with the data in recent experiments.
\end{abstract}
\vspace{4mm}

Bose-Einstein condensates[1-4] of atomic gases offer new opportunities
for studying quantum-degenerate fluids. Almost all the essential properties of
Bose-Einstein condensate systems such as the formation and shape of the condensate and
noncondensate, the nature of its collective excitations and
statistical fluctuations, and the formation and dynamics of vortices
are determined by the strength of  atom-atom interactions. In contrast
with the situation in  traditional superfluids, the strength of 
inter-particle interactions in the atomic condensate can vary over a wide
range of values[5-12]. In particular, the scattering length that
characterizes the atom-atom interactions can be negative, corresponding to an
effective inter-atom attraction.

Most recently, in  trapped atomic
Bose-Einstein condensation, Ketterle's group reported
evidences for 
modifying the scattering length by magnetic-field-induced Feshbach resonance[13]. 
Feshbach resonance have been studied
twenty one years ago[14] at much higher energies, but the Feshbach resonance
energy observed in ultracold atoms can be tuned to near zero.
The theoretical studies of the ultracold atoms with Feshbach resonance[15,16]
show that the two-body interactions responsible for the Feshbach resonance 
produce an additional condensate of molecules[17], which
differs qualitatively from the properties of a single condensate.
In this letter, we discuss the temperature dependance of the hybrid
atomic/molecular condensate and noncondensate density profile as well as
the excitation spectrum.
Especially, for an inhomogeneous system with negative scattering length,
 the excitation spectrum shows a upper constraint on  the atom number,
 which, together with the conventional estimation of the condensate atom
 number, gives the relationship between the atom number and the
 external magnetic field $B$. The theoretical results are in good
 agreement with the experiment.

The binary atom Feshbach resonances studied by varying a strong external
magnetic field in an alkali-atom trap are  hyperfine-induced spin-flip
processes that bring the colliding atoms to a bound molecular state
of different electron spin. This process can be described by
the Hamiltonian
\begin{equation}
H_{FR}=\alpha\int d^3r \Psi_m^+(r)\Psi_a(r)\Psi_a(r)+h.c.,
\end{equation}
where $\Psi_m(r),\Psi_m^+(r)$ ($\Psi_a(r),\Psi_a^+(r)$)
are the annihilation and creation
field operators of the molecules(atoms), $\alpha$ stands for the
transition matrix element proportional to the overlap of the molecular
continuum and bound state wave functions.
Usually, the transition matrix element depends on the magnetic 
field as $\alpha\sim \sqrt{\lambda_a\Delta^2/2|B-B_0|}$,
where $B_0$ is the resonant magnetic field and $\Delta$ characterizes the width of
the resonance as a function of $B$.
The Hamiltonian $H_{FR}$ together with the atomic Hamiltonian
\begin{equation}
H_a=\int d^3r \Psi_a^+(r)[-\frac{\nabla^2}{2M}+V_a(r)-\mu_a]
\Psi_a(r)+\frac{\lambda_a}{2}\int d^3r \Psi_a^+(r)\Psi_a^+(r)
\Psi_a(r)\Psi_a(r),
\end{equation}
the molecular Hamiltonian
\begin{equation}
H_m=\int d^3r \Psi_m^+(r)[-\frac{\nabla^2}{4M}+V_m(r)-\mu_m+\epsilon]
\Psi_m(r)+\frac{\lambda_m}{2}\int d^3r \Psi_m^+(r)\Psi_m^+(r)
\Psi_m(r)\Psi_m(r),
\end{equation}
and the atom-molecule interaction Hamiltonian
\begin{equation}
H_{am}=\lambda\int d^3r \Psi_a^+(r)\Psi_m^+(r)\Psi_m(r)\Psi_a(r)
\end{equation}
forms a total Hamiltonian $H=H_a+H_m+H_{am}+H_{FR}$, which governs the
dynamics of the system under investigation.
Here, $V_{a(m)}(r)$ represents the trapped potential
for atom (molecule), $\mu_{a(m)}$ is the
chemical potential of the atoms(molecules), $\lambda_{a(m)}=\frac
{4\pi a_{a(m)}}{(2)M}$, with $M$ being the atomic mass and $a_{a(m)}$ the
s-wave scattering length of the atom-atom interaction, $\lambda$ denotes the
coupling constant of the atom-molecule interaction, and $\epsilon$
is the energy of the intermediate molecular state relative
to the continuum of the incident atoms.

Separating out the condensate part in the usual fashion
(Bogoliubov approximation) i.e.
\begin{equation}
\Psi_{a(m)}(r)=\phi_{a(m)}(r)+\hat{\varphi}_{a(m)}(r),
\end{equation}
where $\phi_{a(m)}(r)=\langle \Psi_{a(m)}(r)\rangle$
plays the role of a spatially
varying macroscopic Bose field of the atoms(molecules).
The possibility that the resonantly formed quasibound atom pairs
form a molecular condensate was previously suggested by
Timmermans {\it et al.}[15]. Using a Raman photonassociation process, the 
quasibound pairs start to be formed from the atomic condensate and form a 
molecular condensate[17]. Here, we assume that there are
a large number of atoms and molecules in the condensate.
It is easy to show that
the operators $\hat{\varphi}_{a(m)}(r)$ and $\hat{\varphi}_{a(m)}^+(r)$ obey the Bose
commutation relations
\begin{equation}
[\hat{\varphi}_{a(m)}(r),\hat{\varphi}_{a(m)}^+(r^{\prime})]=\delta(r-r^{\prime}).
\end{equation}
In terms of $\hat{\varphi}_{a(m)}$ and $\phi_{a(m)}$,
the Hamiltonian can be expanded as
\begin{eqnarray}
H&=&H_0+H^{'},\nonumber\\
H_0&=&\int d^3r\{ \phi^*_a(-\frac{\nabla^2}{2M}-\mu_a+V_a(r))\phi_a
+\frac{\lambda_a}{2}\phi_a^*\phi_a^*\phi_a\phi_a\nonumber\\
&+&\phi^*_m(-\frac{\nabla^2}{4M}-\mu_m+V_m(r)+\epsilon)\phi_m
+\frac{\lambda_m}{2}\phi_m^*\phi_m^*\phi_m\phi_m\nonumber\\
&+&\lambda\phi_a^*\phi_a\phi_m^*\phi_m+\alpha\phi_m^*\phi_a\phi_a+\alpha
\phi_m\phi^*_a\phi_a^*
\}
\end{eqnarray}
\begin{eqnarray}
H^{'}&=&\int d^3r\{
\hat{\varphi}_a^+(-\frac{\nabla^2}{2M}-\mu_a+V_a(r))\hat{\varphi}_a+2\lambda_a\hat{\varphi}^+_a
\hat{\varphi}_a\phi_a^*\phi_a+\frac{\lambda_a}{2}(\hat{\varphi}_a^+\hat{\varphi}_a^+\phi_a
\phi_a+\hat{\varphi}_a\hat{\varphi}_a\phi_a^*\phi_a^*)\nonumber\\
&+&\hat{\varphi}_m^+(-\frac{\nabla^2}{4M}+\epsilon-\mu_m+V_m(r))\hat{\varphi}_m+2\lambda_m
\hat{\varphi}^+_m
\hat{\varphi}_m\phi_m^*\phi_m+\frac{\lambda_m}{2}(\hat{\varphi}_m^+\hat{\varphi}_m^+\phi_m
\phi_m+\hat{\varphi}_m\hat{\varphi}_m\phi_m^*\phi_m^*)
\nonumber\\
&+&\lambda\hat{\varphi}^+_a\hat{\varphi}_a\phi_m\phi_m^*+\lambda \hat{\varphi}_m^+\hat{\varphi}_m
\phi_a^*\phi_a+\alpha(\hat{\varphi}_a\hat{\varphi}_a\phi_m^*+\hat{\varphi}_a^+\hat{\varphi}_a^+
\phi_m)\}.\nonumber\\
\end{eqnarray}
In derivation of eq.(7,8), the following coupling equations are used[15],
\begin{eqnarray}
&\ &\{-\frac{\nabla^2}{2M}+\lambda_a|\phi_a|^2+V_a(r)+\lambda|\phi_m|^2  \}
\phi_a+2\alpha\phi_m\phi_a^* =\mu_a\phi_a,\nonumber\\
&\ &\{-\frac{\nabla^2}{4M}+\lambda_m|\phi_m|^2+V_m(r)+\epsilon+
\lambda|\phi_a|^2  \}
\phi_m+\alpha\phi_a\phi_a =\mu_m\phi_m.
\end{eqnarray}
This coupling equations may be yielded by the expectation value of the
Heisenberg equations
\begin{equation}
i\hbar\dot{\Psi}_a=[\Psi_a,H], i\hbar\dot{\Psi}_m=[\Psi_m,H],
\end{equation}
and replacing the time derivatives by the chemical potentials
$$i\hbar\dot{\phi}_a\rightarrow\mu_a\phi_a, i\hbar\dot{\phi}_m\rightarrow \mu_m
\phi_m.$$
The chemical potential of the molecules is twice the chemical potential
of the atoms, in accordance with the condition for chemical equilibrium. The
$\alpha-$terms that couple the equations describe tunneling of pairs of atoms
between $\phi_m$ and $\phi_a$ fields, it leads to the form of a
second condensate---molecular condensate in an atomic Bose-Einstein condensate[15-17].
Using the coupling equations(9), Timmermans {\it et al.}[15] investigate the
behaviors of the hybrid atomic/molecular condensates near- and off- resonance.
The Hamiltonian (8) can be diagonalized by using the Bogoliubov transformation
\begin{eqnarray}
\hat{\varphi}_a(r)=\sum_j[u_j(r)\alpha_j-v_j^*(r)\alpha_j^+],\nonumber\\
\hat{\varphi}_a^+(r)=\sum_j[u_j^*(r)\alpha_j^+-v_j(r)\alpha_j],\nonumber\\
\hat{\varphi}_m(r)=\sum_j[x_j(r)\beta_j-y_j^*(r)\beta_j^+],\nonumber\\
\hat{\varphi}_m^+(r)=\sum_j[x_j^*(r)\beta_j^+-y_j(r)\beta_j],
\end{eqnarray}
where the qusiparticle operators $\alpha_j$,$\alpha_j^+$,$\beta_j$,
$\beta_j^+$ obey boson commutation relations
$$[ \alpha_i,\alpha_j^+ ]= \delta_{ij},
 [ \alpha_i,\alpha_j]=[ \alpha_i^+,\alpha_j^+]=0,$$
$$[ \beta_i,\beta_j^+]=\delta_{ij},
 [ \beta_i,\beta_j]=[ \beta_i^+,\beta_j^+]=0,$$
$$[ \alpha_i^+,\beta_j^+]=[ \alpha_i,\beta_j]=[ \alpha_i,\beta_j^+]=0,$$
and $u_j(r),v_j(r),x_j(r),y_j(r)$ are $c$-number functions.
Substituting eq.(11) into eq.(8), one yields
\begin{equation}
H^{'}=\sum_j E_j\alpha_j^+\alpha_j+\sum_i e_i\beta_i^+\beta_i-
\sum_jE_j\int d^3r |v_j(r)|^2-\sum_i e_i\int d^3 r|y_i(r)|^2
\end{equation}
with
\begin{eqnarray}
(-\frac{\nabla^2}{2M}+\lambda|\phi_m|^2+2\lambda_a|\phi_a|^2-\mu_a+V_a(r))
u_j-(\lambda_a\phi_a^*\phi_a^*+2\alpha\phi_m^*)v_j&=&E_ju_j,\nonumber\\
(-\frac{\nabla^2}{2M}+\lambda|\phi_m|^2+2\lambda_a|\phi_a|^2-\mu_a+V_a(r))
v_j-(\lambda_a\phi_a^*\phi_a^*+2\alpha\phi_m^*)u_j&=&-E_jv_j,\nonumber\\
(-\frac{\nabla^2}{4M}+\epsilon+\lambda|\phi_a|^2+2\lambda_m|\phi_m|^2-\mu_m
+V_m(r))
x_j-\lambda_m\phi_m^*\phi_m^*y_j=e_jx_j,\nonumber\\
(-\frac{\nabla^2}{4M}+\epsilon+\lambda|\phi_a|^2+2\lambda_m|\phi_m|^2
-\mu_m+V_m(r))
y_j-\lambda_m\phi_m^*\phi_m^*x_j=-e_jy_j.
\end{eqnarray}
In order to study the temperature dependence of the excitation spectrum
as well as the spatial distribution of the hybrid atom/molecular condensate
and noncondensate, we need to solve the coupled mean-field Bogoliubov equations(13),
and the condensate equation(9) self-consistently. The calculations
procedure can be summarized for an arbitrary confining potential as
follows:
First of all, we solve  eq.(9)  self-consistently, once
$\phi_a$ and $\phi_m$ are known, the solution of $u_j,v_j,x_j$ and $y_j$
can be generated. To illustrate this procedure, we present its first
step of calculations analytically. The trapped potential considered here
is taken to be an isotropic harmonic potential
$V_{a(m)}(r)=\frac 1 2 M\omega_{a(m)}^2r^2$, for which $\phi_a$ and $\phi_m$ are
spherically symmetric functions,
\begin{equation}
\phi_{a(m)}(r)=R_{00}(r)Y_{00}(\theta,\psi),
\end{equation}
with
$$ R_{00}(r)=\alpha^{3/2}\sqrt{\frac{4}{\pi}}exp[-\frac 1 2 \alpha^2 r^2],
Y_{00}(\theta,\psi)=\frac{1}{\sqrt{4\pi}}, \alpha=((2)M\omega)^{1/2}.$$

Rather than solving the coupled equations (13) directly,
we introduce a new method based on the auxiliary functions
\begin{eqnarray}
u_j&=&A_j\langle r|j\rangle_a,v_j=B_j\langle r|j\rangle_a,\nonumber\\
x_j&=&C_j\langle r|j\rangle_m,y_j=D_j\langle r|j\rangle_m,\nonumber\\
\end{eqnarray}
where $|j\rangle_{a(m)}$ is defined by
$$[-\frac{\nabla^2}{2(4)M}+V_{a(m)}(r)]|j\rangle_{a(m)}=\hbar
\omega_{a(m)}(j+\frac 1 2)|j\rangle_{a(m)}.$$
The reason for such selection is that the level
shifts caused by atom-atom interactions weakly depend on the shape of the wave
function. A combination of Eqs.(13-15) gives
\begin{eqnarray}
(\hbar\omega_a(j+\frac 12)+\lambda|\phi_m|^2+2\lambda_a|\phi_a|^2-\mu_a)
A_j-(\lambda_a\phi_a^*\phi_a^*+2\alpha\phi_m^*)B_j&=&E_jA_j,\nonumber\\
(\hbar\omega_a(j+\frac 1 2 )+\lambda|\phi_m|^2+2\lambda_a|\phi_a|^2-\mu_a)
B_j-(\lambda_a\phi_a^*\phi_a^*+2\alpha\phi_m^*)A_j&=&-E_jB_j,\nonumber\\
(\hbar\omega_m(j+\frac 1 2 )
+\epsilon+\lambda|\phi_a|^2+2\lambda_m|\phi_m|^2-\mu_m)
C_j-\lambda_m\phi_m^*\phi_m^*D_j=e_jC_j,\nonumber\\
(\hbar\omega_m(j+\frac 1 2 )+\epsilon
+\lambda|\phi_a|^2+2\lambda_m|\phi_m|^2-\mu_m)
D_j-\lambda_m\phi_m^*\phi_m^*C_j=-e_jD_j,
\end{eqnarray}
the eigenfunctions and the corresponding eigenvalues are given by
\begin{eqnarray}
B_j^{\pm}&=&[\frac {1}{f^{\pm}(r,j)-1}]^{\frac 1 2 },
A_j^{\pm}(r)=f^{\pm}(r,j)B_j^{\pm},\nonumber\\
D_j^{\pm}&=&[\frac {1}{g^{\pm}(r,j)-1}]^{\frac 1 2 },
B_j^{\pm}(r)=g^{\pm}(r,j)D_j^{\pm},
\end{eqnarray}
and
\begin{eqnarray}
E_j^{\pm}(r)&=&\pm\{ (\lambda_a\phi_a^*\phi_a^*+2\alpha\phi_m^*)
-[\hbar\omega_a(j+\frac 1 2 )-\mu_a+\lambda|\phi_m|^2+2\lambda_a|\phi_a|^2]\}
,\nonumber\\
e_j^{\pm}(r)&=&\pm\{ \lambda_m\phi_m^*\phi_m^*-
[\hbar\omega_m(j+\frac 1 2 )-\mu_m+\epsilon+2\lambda_m|\phi_m|^2]\}.
\end{eqnarray}
Here, $$f^{\pm}(r,j)=
\frac{\lambda_a\phi_a^*\phi_a^*+2\alpha\phi_m^*}{\hbar\omega_a(j+\frac 1 2 )
-\mu_a+\lambda|\phi_m|^2+2\lambda_a|\phi_a|^2-(E_j^{\pm})^2},$$
and
$$g^{\pm}(r,j)=\frac{\lambda_m\phi_m^*\phi_m^*}{
\hbar\omega_m(j+\frac 1 2 )
-\mu_m+\epsilon+2\lambda_m|\phi_m|^2-(e_j^{\pm})^2}.$$
These explicit solutions enable us to construct the
one-body density matrix
\begin{eqnarray}
\rho(r,r^{'})&=&\rho_a(r,r^{'})+2\rho_m(r,r^{'}),\nonumber\\
\rho_a(r,r^{'})&=&\phi_a^*(r)\phi_a(r^{'})
\nonumber\\
&+&\sum_{p=\pm,i=1}^{\infty}
[u_i^{p*}(r)u_i^p(r^{'})F_i^p+
v_i^{p*}(r)v_i^p(r^{'})(1+F_i^p)],\nonumber\\
\rho_m(r,r^{'})&=&\phi_m^*(r)\phi_m(r^{'})
\nonumber\\
&+&\sum_{p=\pm,i=1}^{\infty}
[x_i^{p*}(r)x_i^p(r^{'})f_i^p+
y_i^{p*}(r)y_i^p(r^{'})(1+f_i^p)],
\end{eqnarray}
where $F_i^p=\frac{1}{exp(\beta E_i^p)-1}$ and
$f_i^p=\frac{1}{exp(\beta e_i^p)-1}$ are the Bose distribution for the
quasiparticle excitations with energies $E_i^p$ and $e_i^p$, respectively.
Setting $r=r^{'}$, eq.(19) follows the resulting particle density.

We need to point out that  eqs. (17) and (18) are results of the
first step of the numerical calculations. To complete numerical calculations, we
should repeat the above procedures until the eigenvalues $E_j$ and $e_j$
do not depend on position $r$.
In what follows, we present a variation method to study the
excitations at zero temperature. This method was first
introduced in Ref. [18] to study the BEC ground state in the
harmonic trap of boson system, and it was generalized in Ref. [19] to
investigate the excited states in BEC. Considering eq.(13) as well
as
$$\int[u_j(r)u^*_j(r)-v_j(r)v^*_j(r)]dr=1,$$
and
$$\int[x_j(r)x^*_j(r)-y_j(r)y^*_j(r)]dr=1,$$
which were derived from the Bose commutation relation (6) we arrive at
\begin{eqnarray}
E_j&=&\int u_j^*(r)(-\frac{\nabla^2}{2M}+\lambda|\phi_m|^2
+2\lambda_a|\phi_a|^2
+V_a(r))
u_jdr\nonumber\\
&+&\int v_j^*(r)(-\frac{\nabla^2}{2M}+\lambda|\phi_m|^2+2\lambda_a|\phi_a|^2
+V_a(r))v_jdr\nonumber\\
&-&\int u_j^*(r)(\lambda_a\phi_a^*\phi_a^*+2\alpha\phi_m^*)v_j(r)dr
-\int v_j^*(r)(\lambda_a\phi_a^*\phi_a^*+2\alpha\phi_m^*)u_j(r)dr,\nonumber\\
\end{eqnarray}
\begin{eqnarray}
e_j&=&\int x_j^*(r)(-\frac{\nabla^2}{4M}+\varepsilon+\lambda|\phi_a|^2
+2\lambda_m|\phi_a|^2
+V_m(r))x_jdr\nonumber\\
&+&\int y_j^*(r)(-\frac{\nabla^2}{4M}+\lambda|\phi_a|^2+\varepsilon+
2\lambda_m|\phi_m|^2
+V_m(r))y_jdr\nonumber\\
&-&\int x_j^*(r)\lambda_m\phi_m^*\phi_m^*y_j(r)dr
-\int y_j^*(r)\lambda_m\phi_m^*\phi_m^*x_j(r)dr.\nonumber\\
\end{eqnarray}
For simplicity, we study only the case of the spherical harmonic trap. In
this case, we may choose the trial wave functions of the excitation components
$u_j(r),v_j(r)$,$x_j(r)$ and $y_j(r)$ in the form of the spherical harmonic
oscillator wave function $\xi_{n_r,l,m}$ with quantum numbers $(n_r,l,m)$:

\begin{eqnarray}
\left (
\begin{array}{c}
u_j(r)\\
v_j^*(r)\\
\end{array}
\right )
&=&
\left (
\begin{array}{c}
u\\
v^*\\
\end{array}
\right )
\xi_{n_r,l,m}(\omega_{n_rlm},r),
\nonumber\\
\left (
\begin{array}{c}
x_j(r)\\
y_j^*(r)\\
\end{array}
\right )
&=&
\left (
\begin{array}{c}
x\\
y^*\\
\end{array}
\right )
\xi_{n_r,l,m}(\omega_{n_rlm},r),
\end{eqnarray}
where $\omega_{n_rlm}$ is an adjustable scaling factor of variation.
Eqs (20) and (21) show that $E_j$ and $e_j$ take a similar form, hence we here
discuss branches $E_j$ of the excitation spectra in detail.
For $(n_r,l,m)=(0,1,0)$, we have
$$\xi_{0,1,0}=\alpha_{010}^{3/2}[\frac{8}{3\sqrt{\pi}}]^{1/2}\alpha_{010}
re^{-\alpha_{010}^2r^2/2}Y_{1,0}(\theta,\psi).$$
The excitation spectrum in this case is reduced to
\begin{eqnarray}
E&=&E[v,\omega_{010}]=(1+2v^2)[\frac 5 4 \hbar\omega_{010}+\frac 5 4
\hbar\frac{\omega_a^2}{\omega_{010}}]\nonumber\\
&+&\lambda(1+2v^2)N_m\omega_m^{3/2}[\frac{2M}{\pi\hbar}]^{3/2}
[\frac{\omega_{010}}{\omega_{010}+2\omega_m}]^{5/2}\nonumber\\
&+&[2\lambda_a(1+2v^2)-2\lambda_av\sqrt{1+v^2}]N_a\omega_a^{3/2}[\frac
{M}{\pi\hbar}]^{3/2}[\frac{\omega_{010}}{\omega_{010}+\omega_a}]^{5/2}
\nonumber\\
&-&4\alpha v\sqrt{1+v^2}[\frac{\omega_{010}}{\omega_{010}+\omega_m}]^{5/2}
N_m^{1/2}\omega_m^{3/4}[\frac{2M}{\pi\hbar}]^{3/4},
\end{eqnarray}
where $\alpha_{010}^2=\frac{M\omega_{010}}{\hbar}.$Similarly, for
$(n_r,l,m)=(1,0,0),$ we have
$$\xi_{1,0,0}=\alpha_{100}^{3/2}[\frac{8}{3\sqrt{\pi}}]^{1/2}
(\frac 3 2-\alpha_{100}^2r^2)
re^{-\alpha_{100}^2r^2/2}Y_{0,0}(\theta,\psi).$$
And
\begin{eqnarray}
E&=&E[v,\omega_{100}]=(1+2v^2)[\frac 7 4 \hbar\omega_{100}+\frac 7 4
\hbar\frac{\omega_a^2}{\omega_{100}}]\nonumber\\
&+&\lambda(1+2v^2)N_m\omega_m^{3/2}[\frac{2M}{\pi\hbar}]^{3/2}
f(\omega_{100},2\omega_m)\nonumber\\
&+&[2\lambda_a(1+2v^2)-2\lambda_av\sqrt{1+v^2}]N_a\omega_a^{3/2}[\frac
{M}{\pi\hbar}]^{3/2}f(\omega_{100},\omega_a)
\nonumber\\
&-&4\alpha v\sqrt{1+v^2}f(\omega_{100},\omega_m)
N_m^{1/2}\omega_m^{3/4}[\frac{2M}{\pi\hbar}]^{3/4},
\end{eqnarray}
where
$$f(x,y)=\frac 3 2(\frac{x}{x+y})^{3/2}-3(\frac{x}{x+y})^{5/2}+
\frac 5 2(\frac{x}{x+y})^{7/2}.$$
Minimizing the energies of eqs (23) and (24) with respect to the variation
parameters $v$, $\omega_{010}$ and $\omega_{100}$, we can  determine
the excitation spectrum for the mode $(0,1,0)$ and $(1,0,0)$.
The numerical results are illustrated in fig.1 and fig.2.
The dashed lines in figures show the excitation spectrum in atom BEC,
i.e., $\alpha=\lambda=0$. In contrast, the solid line
are those of hybrid atomic/molecular condensates near Feshbach resonance.
From fig.1 and fig.2 we see that while the
excitation frequency for modes $(0,1,0)$ increases due to the Feshbach resonance effect,  
the excitation frequency for the
mode $(1,0,0)$ decreases.
We would like to point out that the numerical results presented here
depend on the coupling constant as well as the parameter $\alpha$.
In Fig.1 and Fig.2, we let $\alpha=5\lambda_a$, and $\lambda_a=0.1$ (arbitrary units).
The other parameters in Fig.1 and Fig.2 are $\omega_m=1.4\omega_a=7500 Hz,$
$N_a=N_m=10^6$.

For
clarity, we illustrate the above somewhat formal discussion by
considering the binary atom system for a uniform system
($V_{a(m)}(r)\rightarrow 0$),
in this case $H_{FR}$ gives a resonant contribution to the atom-atom
interaction strength $a_a$: $a_{eff}=a_0(1+\frac{\Delta}{B_0-B})$,
where $a_0$ is the off-resonant scattering length, and $\Delta$ characterizes
the width of the resonance.
For small $p$, these excitations are phonons, and their energy tends to zero
with $p$. Hence,
\begin{equation}
\mu_a=\lambda|\phi_m|^2+\lambda_a|\phi_a|^2-2\alpha\phi_m^*,
\end{equation}
which leads to
\begin{equation}
E_j^2=E^2(p)=(\frac{\hbar^2}{2m})^2p^2(p^2+16\pi n a_{eff}).
\end{equation}
For a uniform dilute Bose gas with negative scattering length $a_{eff}$,
eq.(26) implys an instability of
those modes with $p^2\leq 16\pi n|a_{eff}|$.
For a gas in a trap, however, the wavenumber cannot be arbitrarily small, and the
minimum value is of order $p_{min}\simeq\pi/R_0$(
$R_0$ is the mean size of the ground state). Hence the system can
remain stable if $\frac{\pi^2}{R_0^2}\geq16\pi n|a_{eff}|.$ Since the
density is of order $n\simeq N/R_0^2$, this means that the critical number
of the system is
\begin{equation}
N_0\simeq \frac{\pi}{16}\frac{R_0}{|a_{eff}|}.
\end{equation}
For a positive scatering length $a_{eff}$, however, there are  not any
constraints in $N$. The Bogoliubov quasiparticle theory shows that the
condensate atoms $N_0$ depends on the scattering length and satisfies
(for $(a_{eff}\frac N V)^{\frac 1 3 }<<1$).
\begin{equation}
N_0=N(1-\frac 8 3 \sqrt{\frac{Na^3_{eff}}{\pi V}}).
\end{equation}
The numerical results of eqs.(27) and (28) are illustrated in Fig.3,
which shows the atom number $N_0$ vs. external magnetic field $B$.
The parameters in Fig.3 are $N/V=N/R^2=10^{15}/cm^3$, $\Delta=0.01mT$.

To sum up, we have derived a set of four coupled  equations 
of the atomic and molecular excitations within standard Hartree-Fock-Bogoliubov 
approximation. As shown in eqs. (9) and (13), the $\alpha$ terms describing the 
process that converts atoms into molecules play an important role in 
atomic/molecular Bose-Einstein condensation.
In particular, two low excitation 
spectrum have been given at zero temperature, which
show that the interaction between the hybrid atomic/molecular BEC increase 
one excitation mode, while they decrease another excitation
mode. The mode (0,1,0) comes from the density fluctuation of the condensate
like vibrating oscillation, in this sense that the mode (0,1,0) increases near the Feshbach resonance
indicates the presence of the Feshbach resonance enhance the density fluctuation like vibrating 
oscillation
in atomic/molecular condensation system, whereas the breath mode(like breathing oscillation) (1,0,0) 
decrease near the Feshbach resonance.
Within the two-body collision regime, we show the atom 
number remained in BEC vs. the external magnetic field $B$, 
the result is in good agreement with the recent experiment.
This work removes from consideration of the case at resonance, since at resonance 
the Bogoliubov approximation is not available( at resonance, there
are few atoms in condensate). The contributions  
of the noncondensate atoms(molecules) to the excitation spectrum
is also ignored (see eq.(9)). These need further investigations.

\ \ \\
\begin{center}
Figure Captions
\end{center}
{\bf Fig.1:} Excitation spectrum of Mode $(0,1,0)$ vs. number of atoms.
Dotted  and dashed line indicate those with and without Feshbach
resonance, respectively.\\
{\bf Fig.2:}Same as fig.1. But for mode $(1,0,0)$.\\
{\bf Fig.3:}The number of atoms in condensate vs. magnetic field $B$.
\end{document}